\documentclass[12pt]{iopart}

\usepackage{graphicx}
\usepackage{subfigure}
\usepackage{amssymb}
\usepackage{cite}
\usepackage[export]{adjustbox}
\usepackage{xcolor}
\usepackage{hyperref}
\usepackage{multirow}

\usepackage{array}

\begin{document}

\title[Influence of field of view in visual prostheses design]{Influence of field of view in visual prostheses design: Analysis with a VR system}

\author{Melani Sanchez-Garcia*, Ruben Martinez-Cantin, Jesus Bermudez-Cameo, Jose J. Guerrero}

\address{Instituto de Investigación en Ingeniería de Aragón, (I3A). Universidad de Zaragoza, Spain}
\ead{mesangar@unizar.es}
\vspace{10pt}

\begin{abstract}

Visual prostheses are designed to restore partial functional vision in patients with total vision loss. Retinal visual prostheses provide limited capabilities as a result of low resolution, limited field of view and poor dynamic range. Understanding the influence of these parameters in the perception results can guide prostheses research and design. In this work, we evaluate the influence of field of view with respect to spatial resolution in visual prostheses, measuring the accuracy and response time in a search and recognition task. Twenty-four normally sighted participants were asked to find and recognize usual objects, such as furniture and home appliance in indoor room scenes. For the experiment, we use a new simulated prosthetic vision system that allows simple and effective experimentation. Our system uses a virtual-reality environment based on panoramic scenes. The simulator employs a head-mounted display which allows users to feel immersed in the scene by perceiving the entire scene all around. Our experiments use public image datasets and a commercial head-mounted display. We have also released the virtual-reality software for replicating and extending the experimentation. Results show that the accuracy and response time decrease when the field of view is increased. Furthermore, performance appears to be correlated with the angular resolution, but showing a diminishing return even with a resolution of less than 2.3 phosphenes per degree. Our results seem to indicate that, for the design of retinal prostheses, it is better to concentrate the phosphenes in a small area, to maximize the angular resolution, even if that implies sacrificing field of view.

\end{abstract}

\vspace{2pc}

%
%
%

\section{Introduction}

\begin{figure}[t!]
\centering
\includegraphics[width=1\textwidth]{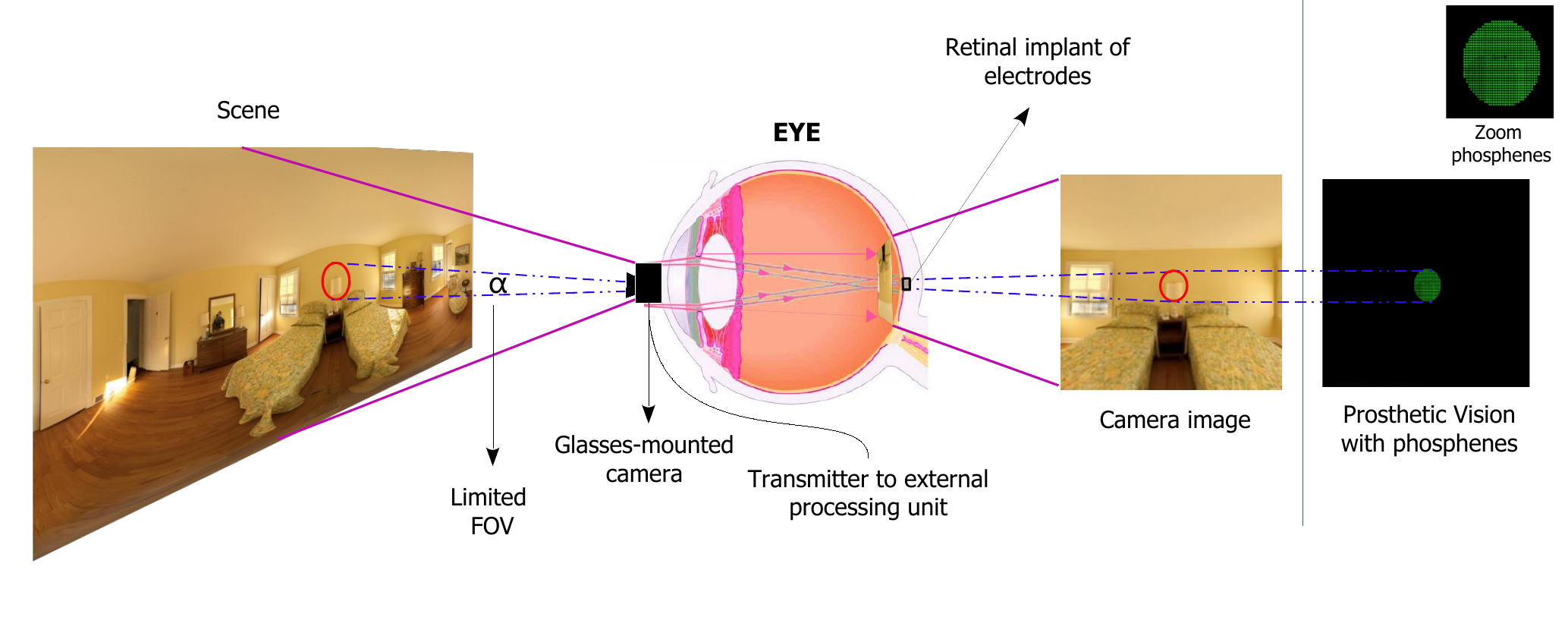}
\caption{\textbf{Overview of a retinal prosthesis.} The external and internal components include a glasses-mounted camera, an external processing unit and an implanted electrode array. First, the external camera acquires an image. Then, the external processor converts the image to a suitable pattern of electrical stimulation of the retina through an electrode array. The result is a phosphene image with limited field of view (FOV).}
\label{fig1}
\end{figure}

Retinitis pigmentosa and age-related macular degeneration are two important causes of visual impairment and blindness \cite{hartong2006retinitis, yu2005retinal}. These diseases involve gradual loss of photoreceptor, or rod cells, while generally preserving the inner retinal cells. This results in a progressive loss of vision. Retinal prostheses are a promising technology to improve vision in patients with such advanced degenerative diseases \cite{zhou2013argus,cheng2017advances,lovell2007advances,luo2016argus}. These visual prostheses can partially restore vision, bypassing damaged photoreceptors and electrically stimulating the surviving retinal cells, such as the retinal ganglion cells \cite{zapf2015towards}. The Argus II epiretinal prostheses (Second Sight Medical Products Inc., Sylmar, CA, USA) is the most widely used retinal prostheses world-wide \cite{luo2016argus,zhou2013argus}. It is made up of an external component (glasses-mounted camera) and an implanted component (electrode implant), as can be seen in Figure~\ref{fig1}. The camera acquires images from the real-world that are transmitted to a portable visual processing unit linked to the camera. The processed information is sent to the retina via electrical impulses in the implant by an electrode array. The stimulation can activate a group of neurons in a small localized area of the retina leading to percepts of spots of light known as “phosphenes” \cite{humayun2012interim}. The brain interprets patterns of phosphenes in the restricted area as visual information. Results in implanted subjects have demonstrated partial visual restoration, with improvement in both coarse objective function and performance of everyday tasks \cite{da2016five,humayun2009preliminary,dorn2013detection,ahuja2011blind,ho2015long,da2013argus,luo2015use,dagnelie2017performance,luo2016long,kotecha2014argus}.

There are still physiological and technological limitations of the information received by implanted patients. The number of electrodes and implant size limit the maximum amount of information that can be provided by the stimulating array. This fact has restricted the degree of visual resolution (up to 1500 phosphenes) and dynamic range of the visual perception (8 gray levels) that can be delivered to the user. Depth perception is also not possible due to low resolution or monocular implantation. Although most current implants have shown good results using static electrode stimulation, they only elicit perception of multiple isolated phosphenes, sometimes resulting in a combination of non-coherent shapes. Therefore, alternative approaches using dynamic activation of an electrode sequence are being studied \cite{beauchamp2020dynamic}. Additionally, field of view (FOV) is a key limitation affecting visual experience of recipients. The ocular anatomy and surgery are two major limiting factors for the narrow FOV \cite{yue2016retinal, li2018optimized}. Current systems provide a FOV of approximately 18$^{\circ}$ $\times$ 11$^{\circ}$ in the retinal area, which correspond to the FOV subtended by the electrode implant on the retina, as can be seen in Figure~\ref{fig1}.

The constrained FOV limits mobility and recognition capabilities of the prosthetic vision system reducing quality of life\cite{barhorst2016effects, haymes1996mobility, kuyk1998visual}. To improve the design of current retinal prostheses in terms of FOV and resolution, one of the questions to be answered is whether expanding the FOV of the input image can benefit implanted subjects. Some studies have attempted to improve the narrow FOV on the clever design of microelectrode arrays \cite{lohmann2019very,ameri2009toward}. Lohmann et al. \cite{lohmann2019very} developed a flexible and thin retinal implant consisting of a multielectrode array approximately three times the size of the comparable epiretinal Argus II device which shows the possible recovery of meaningful peripheral vision. Other devices for retinitis pigmentosa patients (tunnel vision) have been designed to expand the projected FOV on their retina, for example, minimizing the scene zooming out using a camera \cite{alshaghthrah2014study,kennedy1977field}. He et al. \cite{he2019trade} studied the influence of FOV in the Argus II system with thermal imaging by changing the mapping (zoom out) between the sensor and the electrode array. Zooming out turns out in a larger visual angle mapped onto the same implant region on the retina, effectively decreasing the spatial resolution of their prosthetic vision. Alternatively, Ameri et al. \cite{ameri2009toward} have designed an implant with wider FOV, by spreading the electrodes in a wider retinal surface. This setup seem to have advantages for motion perception and head scanning, at the expense of a reduced electrode density and fine detail. Therefore, the optimal implant area for a fixed number of electrodes remains an open question, which we address in this study.

One of the concerns that has limited the development and wider use of retinal prosthetic devices is how to evaluate their utility and function in terms of benefit to the patient and, consequently, how to predict in which direction to develop these devices \cite{bloch2019advances}. Most studies in this field have incorporated performance-based measures and questionnaires to try to understand the relative importance of visual parameters such as resolution, FOV and visual acuity, in the performance of daily tasks in subjects with visual impairment. However, studies with implanted subjects have limited statistical power and require cumbersome experiments. For example, the study of He et al. \cite{he2019trade} was limited to four implanted participants. Simulated prosthetic vision (SPV) opens the opportunity to evaluate potential and forthcoming functionality, in early stages of design, of these implants with larger studies by using normally sighted participants. SPV is a standard procedure for non-invasive evaluation using participants without visual impairments. Furthermore, SPV systems allow for quick or even interactive modifications of the parameters of the simulation.

Researchers have previously used SPV for analysis of the visual perception in terms of resolution or FOV with normally sighted subjects. For example, Fornos et al. \cite{fornos2008simulation} used SPV to study how such restrictions of the amount of visual information provided would affect performance on simple pointing and manipulation tasks. Li et al. \cite{li2018image} applied an image processing algorithm to the image-to-electrode mapping process which improved the ability of the prosthesis visual perception under SPV. Hayes et al. \cite{hayes2003visually} designed a set of tasks to assess performance of object recognition and manipulation and reading using different sizes of electrode array. Contrary to our approach in this paper, they used a constant electrode density, meaning that wider arrays also had larger number of electrodes. In a subsequent study, Dagnelie et al. \cite{dagnelie2007real} explored minimal visual resolution requirements of a simulated retinal electrode array for mobility in real and virtual environments, experienced by normally sighted subjects in video headsets.

These experiments relied on some form of head-mounted device which allowed a virtual-reality experience during the experiment, but the \emph{immersiveness} of experience was poor due to the technological limitations compared to modern commercial VR systems. Furthermore, most SPV studies use computer screens to present phosphenic images \cite{vurro2014simulation, sanchez2020semantic,fornos2005simulation}. 
Modern commercial VR systems have specifications in terms of resolution, latency and response time that allows a fully immersive experience with unexpensive equipment and computers. Using these devices, subjects are immersed and able to interact with complex environments. SPV with commercial VR systems, such as the Oculus Rift used in this work, can be a useful tool to evaluate everyday tasks in more realistic setups and complex simulations. Furthermore, prosthetic vision can be assessed in controlled, real or simulated environments.

In this work, we evaluated the influence of FOV and resolution of the prostheses on the subject's performance in a recognition task, since it is of high priority for patients with visual diseases such as retinitis pigmentosa. Concretely, we analyzed an object search and recognition task performance in indoor scenes with different reduced FOVs and resolutions limited to hundreds of electrodes. For that, we present a new VR system for more realistic SPV environments using panoramic scenes. The VR system could be extended to more complex tasks because it supports realistic environments. This system acts as an electronic visual aid that attaches to the user’s head and presents information directly to the user’s eyes. The panoramic scenes (360 degrees) are intended to allow subjects to feel immersed in the scene.

\begin{figure}[t!]
\centering
\includegraphics[width = 6in]{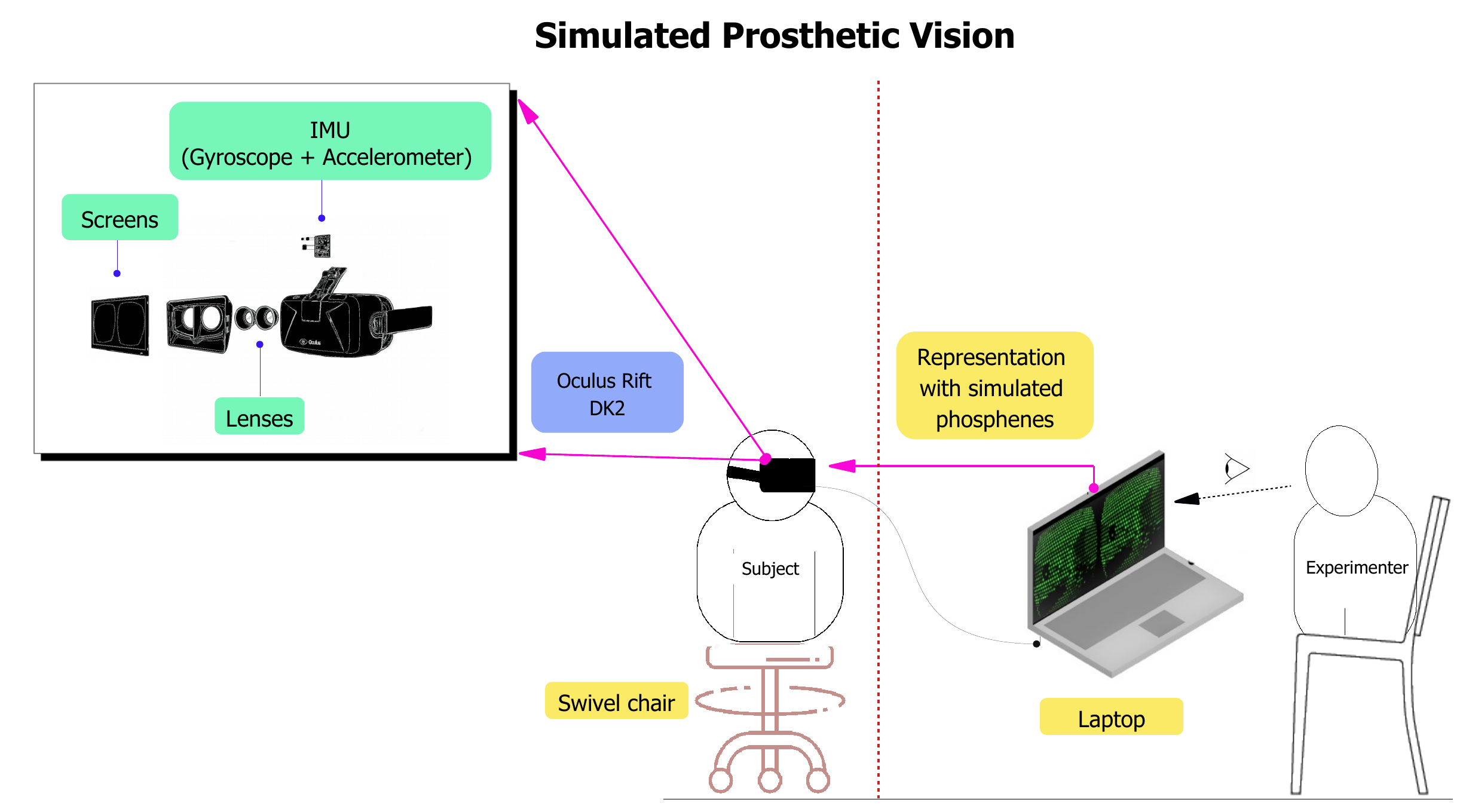}
\caption{\textbf{SPV system}. The components consist of an Oculus Rift powered by a consumer level laptop. The VR system is composed by two lenses, two screens and a suite of internal sensors (gyroscope and accelerometer). The representation with simulated phosphenes is displayed on the laptop screen as well as on the Oculus system worn by the subjects. During the experiment, subjects were seated in a swivel chair allowing them to scan the entire scene all around them (360 degrees).}
\label{fig2}
\end{figure}

\section{Materials and Methods}
We evaluate the influence of FOV on object search and recognition tasks using SPV through a VR system. The SPV system is a standard procedure for non-invasive evaluation using normal vision subjects. This methodology allows controlled evaluation of normally sighted subject response and task performance which is fundamental to know the way humans perceive and interpret phosphenized renderings. SPV also offers the advantage of adapting implant designs to improve the perceptual quality without involving implanted subjects.

\subsection{Participants}
Twenty four subjects aged 20-30 (6 female, 18 male) participated in the experiment. They had no visual problems or wore their normal optical correction during the experiment.

\subsection*{Ethical statement}

The research process was conducted according to the ethical recommendations of the Declaration of Helsinki. The research protocol used for this study is non-invasive, purely observational, with absolutely no-risk for any participant. There was no personal data collection or treatment and all subjects were volunteers. Subjects gave their informed written consent after explanation of the purpose of the study and possible consequences. The consent allowed the abandonment of the study at any time. All data were analyzed anonymously. The experiment was approved by the Aragon Autonomous Community Research Ethics Committee (CEICA).

\begin{figure}[t!]
\centering
\includegraphics[width = 5in]{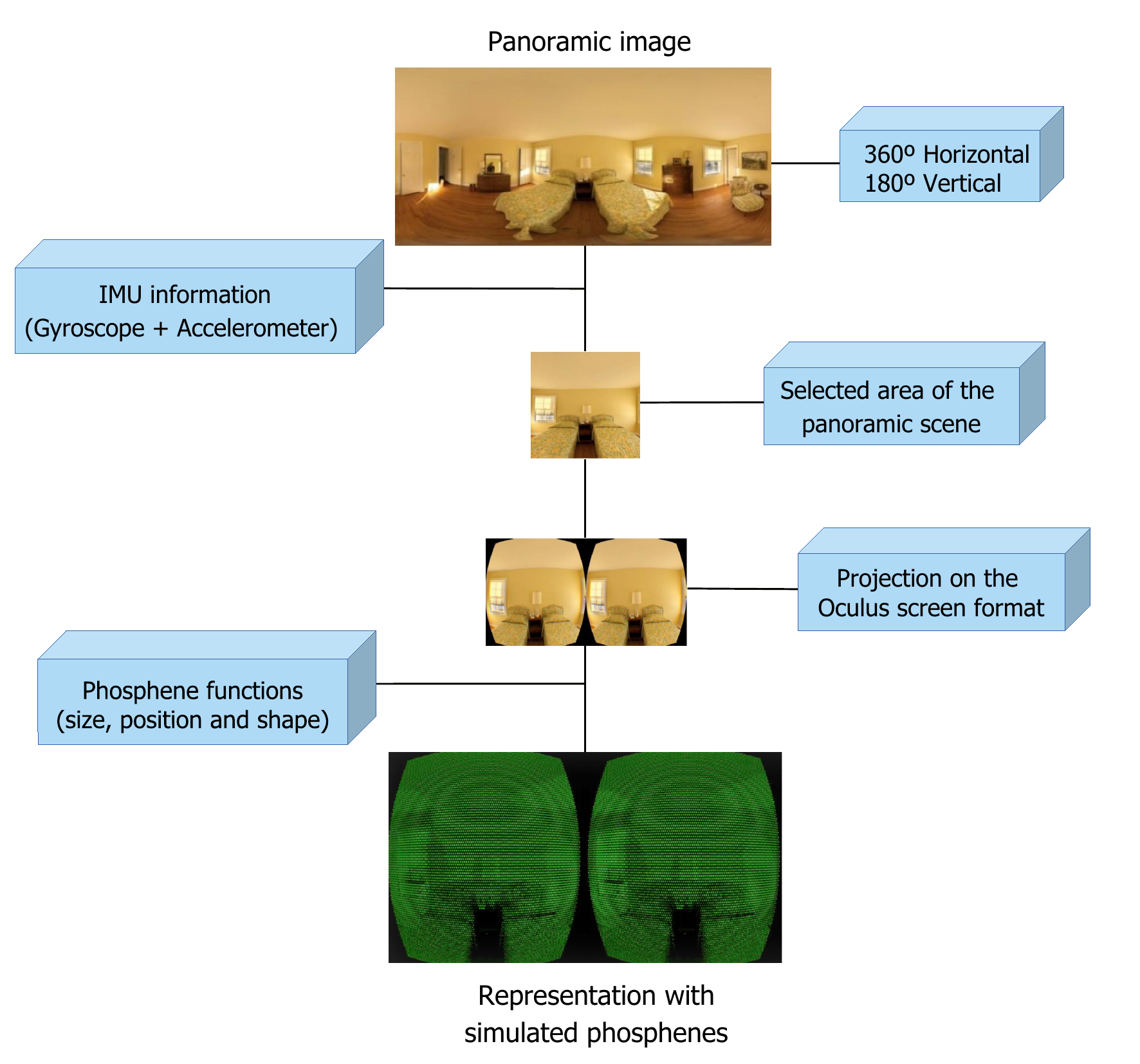}
\caption{\textbf{Data process}. The input scene in our VR system is a panoramic image in equirectangular representation. The system estimate the head orientation using the IMU sensors (gyroscope and accelerometer) allowing to choose the area of the panoramic scene that is being observed at the moment. The area selected is then projected on the two Oculus screens and represented using simulated phosphenes.}
\label{fig3}
\end{figure}

\begin{figure}[t!]
\centering

\begin{tabular}{@{} c @{}}
Oculus \\ FOV \\[1ex]
\raisebox{0.5in}{\rotatebox[origin=c]{90}{60 degrees}}\hspace{0.1em}
\raisebox{0.6in}{\rotatebox[origin=c]{90}{}}\hspace{0.3em}
\includegraphics[width = 1.4in]{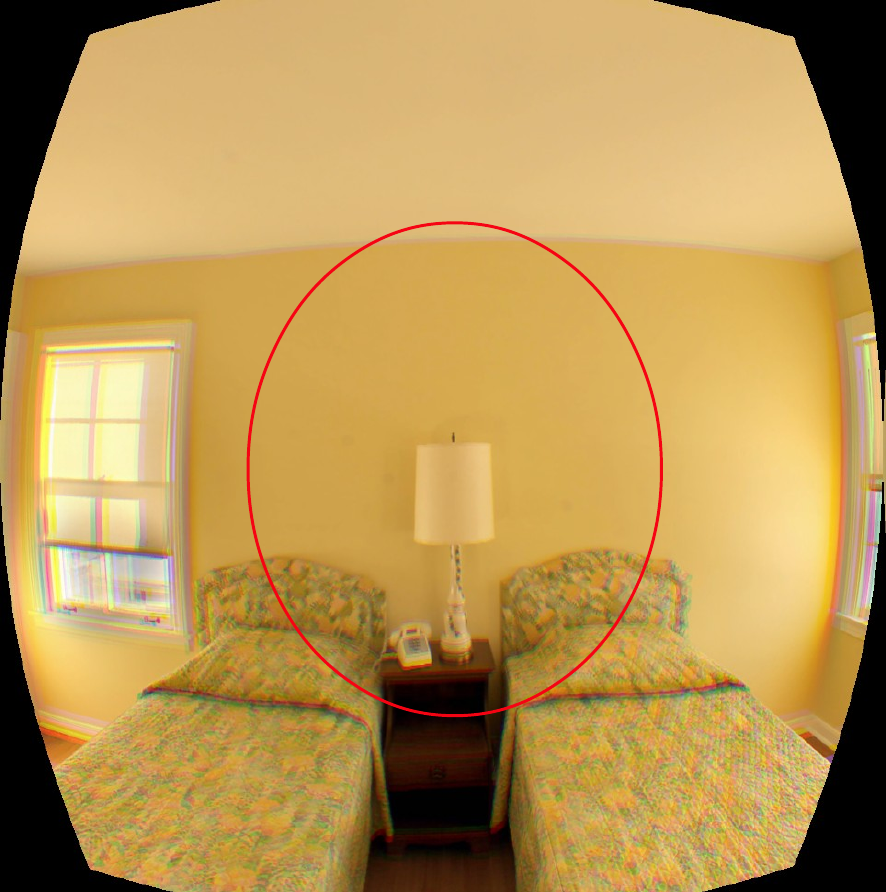}
  \end{tabular}
\begin{tabular}{@{} c @{}}
    Reduced\\ FOV \\[1ex]
\includegraphics[width = 1.4in]{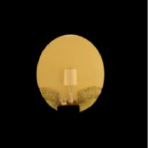}
  \end{tabular}
\begin{tabular}{@{} c @{}}
   200 \\phosphenes \\[1ex]
\includegraphics[width = 1.4in]{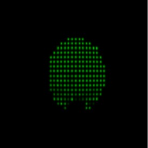}
  \end{tabular}
\begin{tabular}{@{} c @{}}
  500 \\phosphenes \\[1ex]
\includegraphics[width = 1.4in]{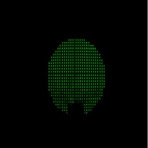}
\end{tabular}

\begin{tabular}{@{} c @{}}
\raisebox{0.5in}{\rotatebox[origin=c]{90}{40 degrees}}\hspace{0.1em}
\raisebox{0.6in}{\rotatebox[origin=c]{90}{}}\hspace{0.3em}
\includegraphics[width = 1.4in]{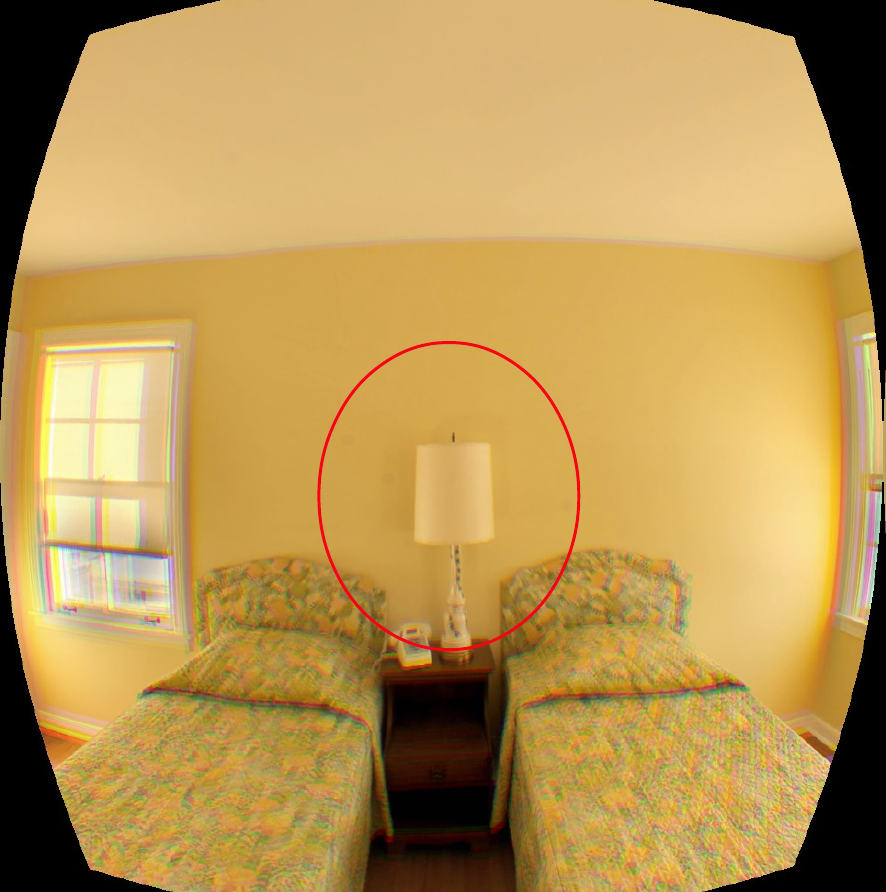}
  \end{tabular}
\begin{tabular}{@{} c @{}}
\includegraphics[width = 1.4in]{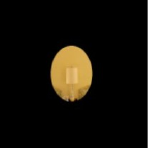}
  \end{tabular}
\begin{tabular}{@{} c @{}}
\includegraphics[width = 1.4in]{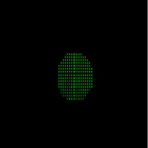}
\end{tabular}
\begin{tabular}{@{} c @{}}
\includegraphics[width = 1.4in]{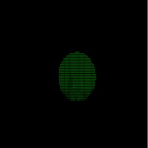}
\end{tabular}

\begin{tabular}{@{} c @{}}
\raisebox{0.5in}{\rotatebox[origin=c]{90}{20 degrees}}\hspace{0.1em}
\raisebox{0.6in}{\rotatebox[origin=c]{90}{}}\hspace{0.3em}
\includegraphics[width = 1.4in]{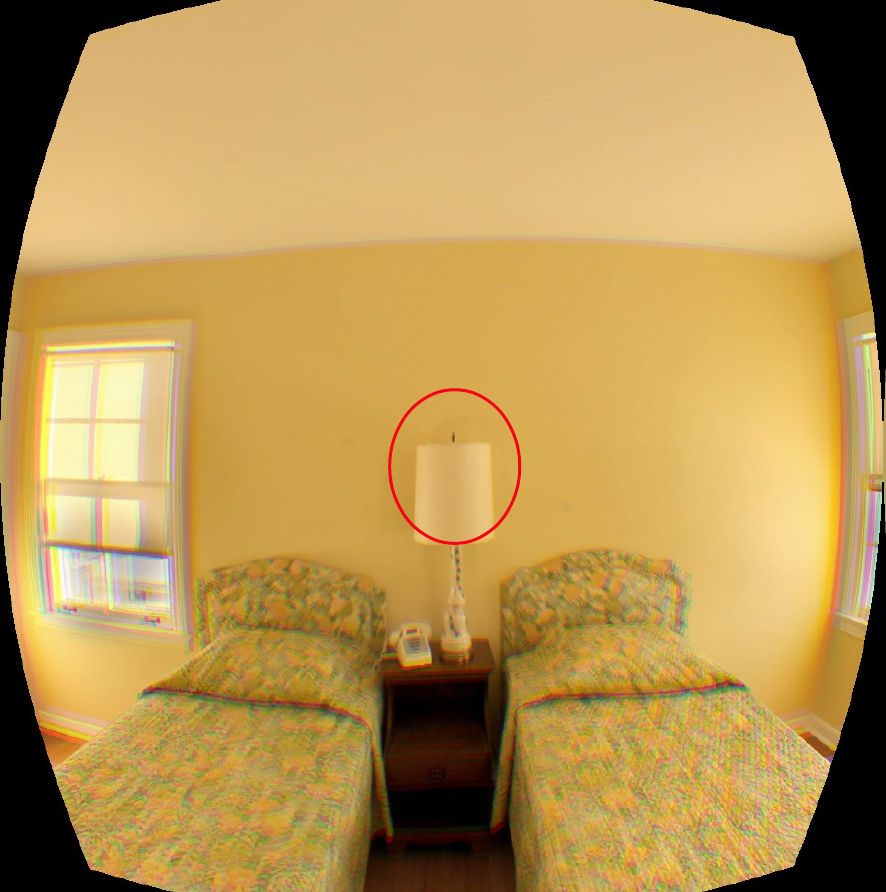}
  \end{tabular}
\begin{tabular}{@{} c @{}}
\includegraphics[width = 1.4in]{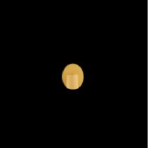}
  \end{tabular}
\begin{tabular}{@{} c @{}}
\includegraphics[width = 1.4in]{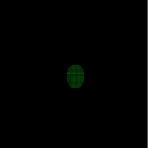}
\end{tabular}
\begin{tabular}{@{} c @{}}
\includegraphics[width = 1.4in]{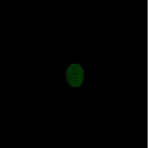}
\end{tabular}

\caption{\textbf{Stimuli conditions in the experiment.} Rows: different circular FOVs used in the experiment (60, 40 and 20 degrees). Columns: different resolutions used in the experiment (200 and 500 phosphenes). Note that in the last row only part of the lamp is visible, the beds cannot be seen without moving the point of view.}
\label{fig4}
\end{figure}


\subsection{Simulated prosthetic vision (SPV)}

This section describes the SPV system including the hardware specifications, software components and phosphene generation.

\subsubsection{Hardware.}\label{section:hardware}As shown in Figure~\ref{fig2}, the experiment was conducted on an Oculus Rift powered by a consumer level laptop (Intel(R) Core(TM) i5-8265U CPU). This system is capable of working in real time with any mid-range laptop. The VR system, Oculus Rift DK2, is composed by two lenses, two screens and an inertial measurement unit (IMU) with gyroscopes and accelerometers, a standard setup for most commercial VR systems. It contains 5.7 inch dual OLED screens with a resolution of 960 x 1080 projected on to each eye. Each display is projected into the eye using a lens with pincushion distortion to provide peripheral vision. In our experiments we mostly use the central part of the display which remains undistorted. The representation with simulated phosphenes was displayed on the VR system worn by the participants as well as on the laptop screen for the experimenter to check the progress. For the head mounted display, we use a single channel (green) to avoid the chromatic aberration of the device lenses. During the experiment, participants were seated in a swivel chair allowing them to scan the entire scene with head rotation movements.

\subsubsection{Software.}\label{section:software}The implementation was done in C++, using the Oculus Rift SDK to connect with the VR system and OpenCV for image processing. Our software is compatible with the Windows operating system. Figure~\ref{fig3} shows the data process designed to generate the stimuli for the VR system. Starting from a panoramic scene capturing 360$^\circ$ of horizontal FOV and 180$^\circ$ of vertical FOV, the system estimates the orientation of the user using the information collected from the IMU (Figure~\ref{fig2}). The selected area is then projected on the two Oculus displays and finally represented with simulated phosphenes. The projection models for the VR system can be found in the \textit{Appendix A}. 

Our phosphene map configuration is similar to the frameworks of McKone et al. \cite{mckone2018caricaturing} and Chen et al. \cite{chen2009simulating}. We approximate the phosphenes as circular dots with a Gaussian luminance profile --each phosphene has maximum intensity at the center and gradually decays to the periphery, following a Gaussian function--. The intensity of a phosphene is directly extracted from the intensity of the same region in the image. For our experiments, each phosphene has 8 intensity levels. The size and brightness are directly proportional to the quantified sampled pixel intensities. The phosphene map is calculated and updated with respect to head orientation in real time. The complete process of phosphene generation can be found in the \textit{Appendix B}.

The software used in the experiment to simulate the prosthetic vision with the VR system is available at \url{http://webdiis.unizar.es/~rmcantin/index.php/Research/Vrfov}.

\subsection{Procedure}

\begin{figure}[t!]
\centering
\includegraphics[width = 6.3in]{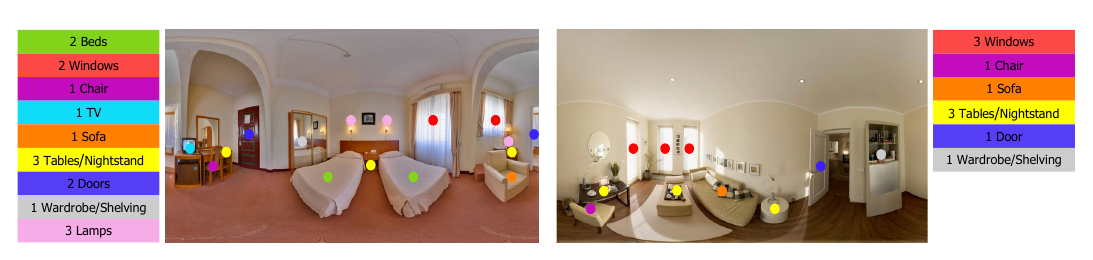}
\caption{\textbf{Object classes considered during the experiment}. Subjects have to recognize the main objects of the scene, those objects that are usually present in hotel rooms such as bed, window, chair, tv, sofa, table/nightstand, door, wardrobe/shelving and lamp.}
\label{fig5}
\end{figure}

The experiment was conducted using a selection of indoor panoramic scenes from a public database of Xiao et al. \cite{xiao2012recognizing} that are adapted to our SPV system. The resolution of the panoramic scenes is 1024 x 512 pixels. All the scenes from the database are hotel rooms containing objects such as beds, tables, chairs, windows and doors, among others (see Figure~\ref{fig5}). We removed several scenes because they had external distractions, such as signatures and watermarks, that could affect the experiment. From them, we randomly selected 50 scenes. The scenes were presented to the subjects using different stimuli conditions based on two resolutions (200 and 500 phosphenes) and three circular FOVs (20, 40 and 60 degrees), as can be seen in Figure~\ref{fig4}. We used a circular FOV to avoid directionality in the searching process that might bias the results. We selected these particular resolutions and FOVs based on current retinal prostheses \cite{bloch2019advances,endo2019influence}, although our VR platform allows to quickly change those parameters. The total number of stimuli scenes generated for the experiment was 300.

For the formal experiment, participants were recruited to complete an object search and recognition task using the SPV system. By turning the swivel chair and head, subjects had to scan the entire scene by changing the head orientation, but not the position. At the same time, participants had to search and recognize the main objects of the scene such as \emph{bed, window, chair, tv, sofa, table/nighstand, door, wardrobe/shelving and lamp} (see Figure~\ref{fig5}). Small objects such as telephones, vases or remote controls were not taken into account. We took into account both the type of recognized object and the number of each of them. At the beginning of the experiment, subjects were trained during 2 minutes on a test set of images. Participants were informed that all scenes were indoor scenes, but they were not informed about the class and number of objects in each scene nor the different stimuli conditions. The demo images were not included in the experiment to avoid learning effects.

\begin{figure}[t!]
\centering
\includegraphics[width = 6in]{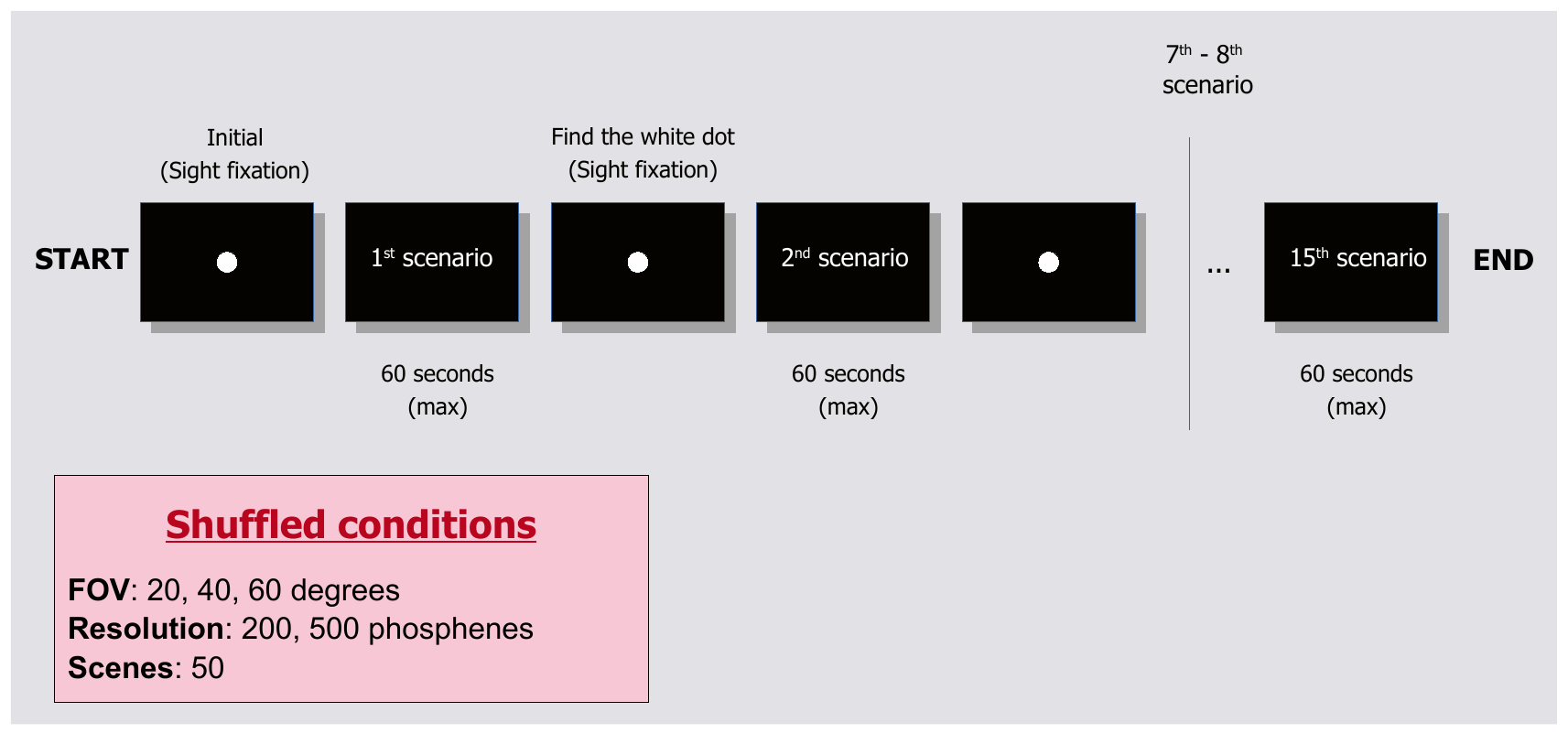}
\caption{\textbf{Trial setup}. To generate the scene in each step of the trial sequence we used shuffled conditions of FOV, resolution and scenes. During the experiment, each scene appeared for 60 seconds and switched for the next scene automatically. Break time in the middle of the sequence (between $7^{th}-8^{th}$ scene) was 60 seconds. The complete experiment took approximately 20 minutes. }
\label{fig6}
\end{figure}

Figure~\ref{fig6} shows the trial setup. Each trial consisted of a sequence of fifteen scenes generated by choosing a resolution, a FOV and a scene, among the shuffled conditions. Initially, a white dot was displayed in the center of the screen indicating where the participants had to maintain fixation until the beginning of the task. Next, each scene was presented and the user has a maximum of 60 seconds to scan the scene moving the chair and their head. The following scene was displayed when the participant verbally ordered that the task had been completed (``I cannot see more objects") or after 60 seconds. This procedure was repeated for each scenario of the trial. Between two scenes the participants were instructed to ``look for" the white dot to fixate the gaze in a steady position before the next scene. There was a break time in the middle of the sequence (between $7^{th}-8^{th}$ scene) of 60 seconds.

The participants verbally indicated the type of objects seen in each scene during the scanning to minimize distractions. The responses of each stimulus scene were timed recorded and annotated by the experimenter. If the participants did not respond within the 60 seconds frame, the result of that scene was considered as ``no object was recognized'' and a time of 60 seconds was recorded. Participants did not get feedback of their responses. The complete experiment took approximately 20 minutes per participant.

\subsection{Statistical analysis}

Data were analyzed using two-way ANOVA and post hoc-test with Tukey's method to evaluate simultaneously the effect of the two grouping variables (resolution and FOV) on the response variables object recognition and recognition time with $p=$ 0.05, * $<$ 0.05 , ** $<$ 0.01, *** $<$ 0.001 and \emph{ns} not significant.

\section{Results}
\label{sec:results}

\begin{figure}[t!]
\centering
\subfigure[]{\includegraphics[width = 3in]{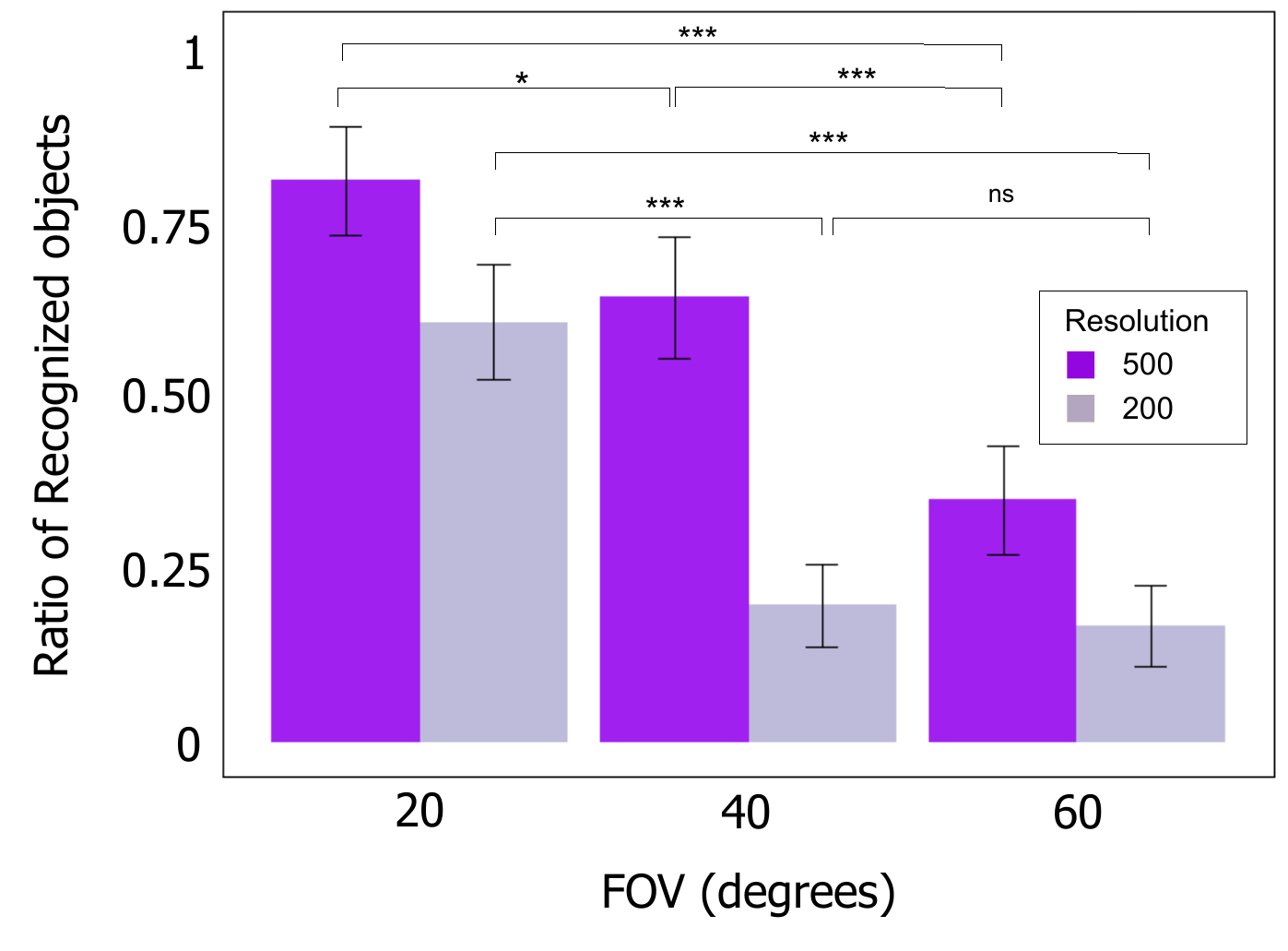}\label{fig7a}}
\subfigure[]{\includegraphics[width = 3in]{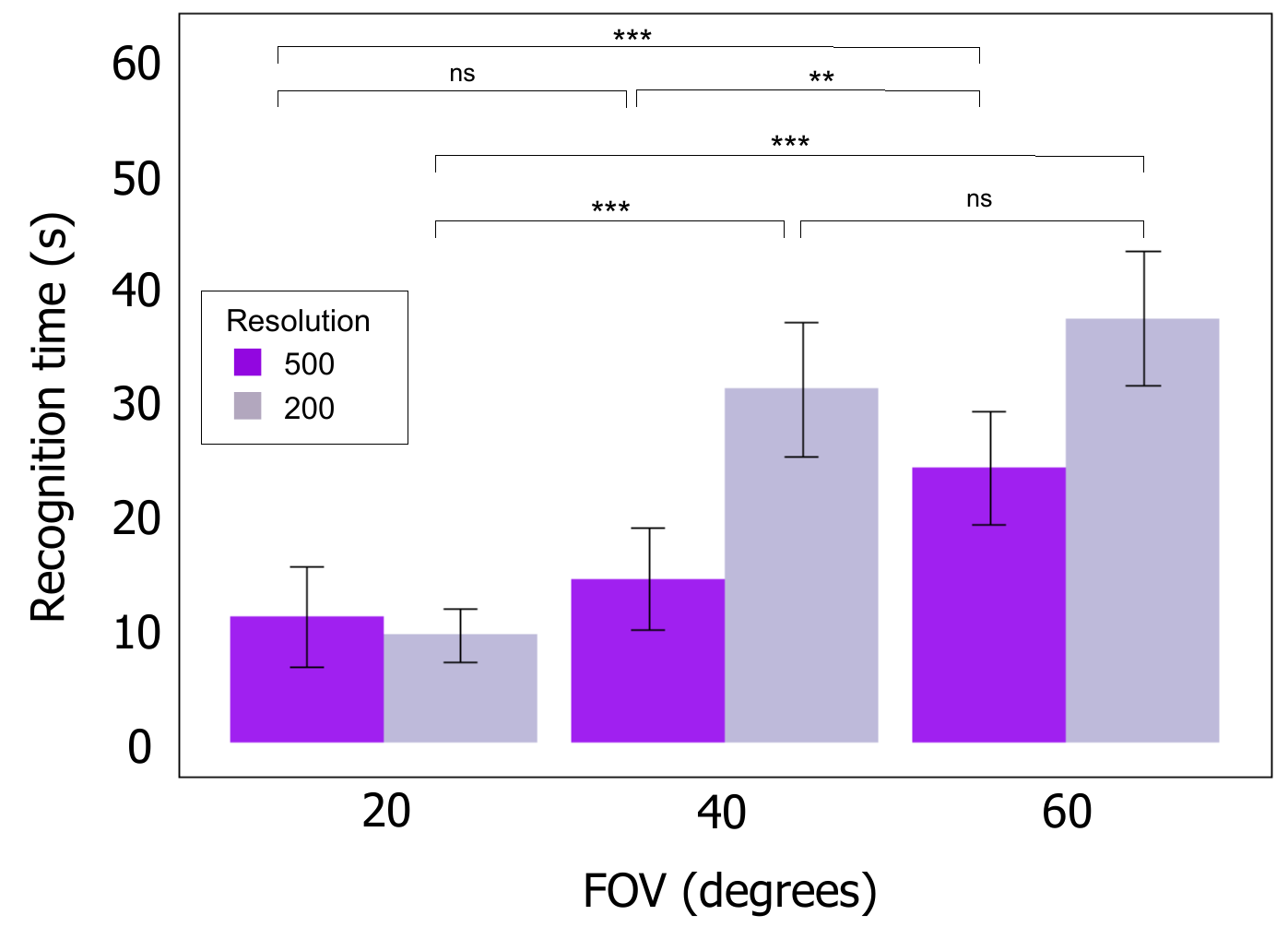}\label{fig7b}}
\caption{Ratio of recognized objects and recognition time. \subref{fig7a} Object recognition results for 20, 40 and 60 FOVs and for the two resolutions, 200 and 500 phosphenes. High scores indicate that subjects were able to recognize most of the objects in each scene. \subref{fig7b} Recognition time results for 20, 40 and 60 FOVs and for the two resolutions, 200 and 500 phosphenes. High scores indicate that the subjects needed more time to perform the recognition task. ***=p$<$.001; **=p$<$.01; *=p$<$.05; ns=p$>$.05.}
\label{fig7}
\end{figure}

The performance for all resolutions and FOVs is summarized in Figure~\ref{fig7}. The results show the ratio of recognized object and recognition time (mean $\pm$ standard deviation) for aggregated data from all subjects and all images. For the same scene, we normalized the results of the object recognition with all the conditions of the experiment, with ``1'' being the experiment case with the highest number of recognized objects for a particular scene. The time recorded in each scene was the time it took the subject to recognize the first object. We also performed a test to determine if the mean difference between specific pairs of conditions are statistically significant using Tukey's method with a significant level $\alpha = 0.05$.

Figure~\ref{fig7}\subref{fig7a} shows the object recognition performance for the three FOVs (20, 40 and 60 degrees) and the two resolutions (200 and 500 phosphenes). For the same resolution, the object recognition performance decreases as the FOV increases. For the resolution of 500 phosphenes the average performance is 80.66$\pm$7.85, 63.84$\pm$8.68 and 34.74$\pm$7.83 for 20, 40 and 60 degrees respectively. For the resolution of 200 phosphenes the average performance is 60.24$\pm$8.28, 19.65$\pm$5.97, 16.58$\pm$5.82 for 20, 40 and 60 degrees respectively. No significant differences were found for 40-60 FOVs for 200 phosphenes (p=0.8056), as performance was very poor in both cases. Comparing the performance for the same FOV, the performance increases with the number of phosphenes for all cases.

Figure~\ref{fig7}\subref{fig7b} shows the recognition time for the three FOVs and two resolutions. For the same resolution, the recognition time increases as the FOV increases. For the resolution of 500 phosphenes the average recognition time is 10.81$\pm$4.32s, 14.10$\pm$4.41s and 23.71$\pm$4.91s for 20, 40 and 60 degrees respectively. For the resolution of 200 phosphenes the average recognition time is 9.24$\pm$2.35s, 30.50$\pm$5.79s, 36.61$\pm$5.77s for 20, 40 and 60 degrees respectively. No significant differences were found for 20-40 FOVs for 500 phosphenes (p=0.733) and similarly to 40-60 FOVs for 200 resolution (p=0.199). Comparing the performance for the same FOV, the recognition time decreases with the number of phosphenes.
Table~\ref{tab:table1} shows the number of cases considered failure, where no object has been recognized in 60s. The condition 60 FOV and 200 resolution had more number of failures. Contrary, the condition 20 FOV and 200 resolution had fewer number of failures.

\begin{figure}[t!]
\centering
\subfigure[]{\includegraphics[width = 3in]{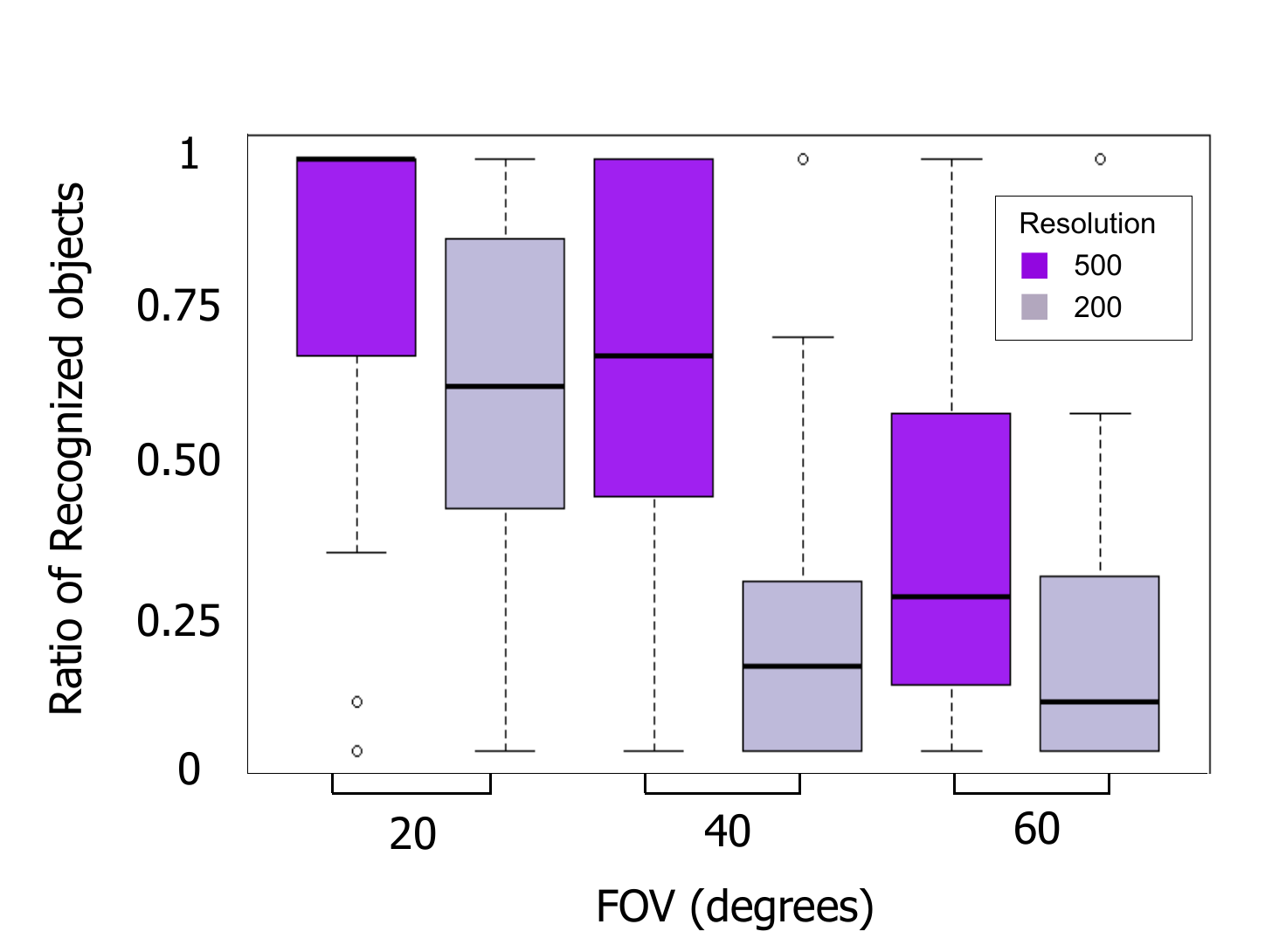}\label{fig8a}}
\subfigure[]{\includegraphics[width = 3in]{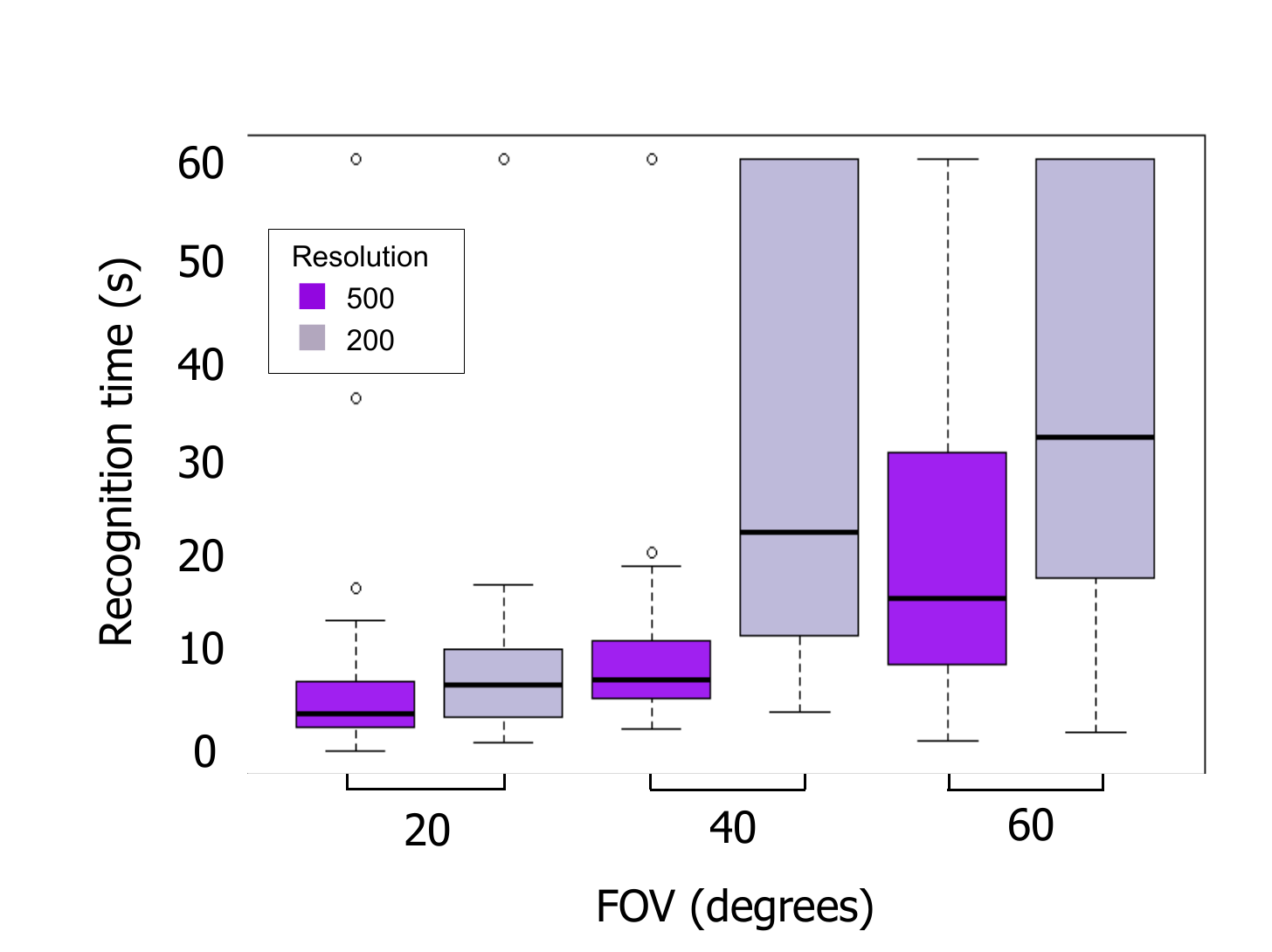}\label{fig8b}}
\caption{Ratio of recognized objects and recognition time. a) Box-plot for object recognition and b) box-plot for recognition time.}
\label{fig8}
\end{figure}

Figure~\ref{fig8} shows box-plots of data distribution with $25$, $50$ and $75th$ quartiles for recognized objects and recognition time. In the recognized object (see Figure~\ref{fig8}\subref{fig8a}), there is difference between the two resolution (200 and 500 phosphenes) for 20 and 40 FOVs. On the other hand, for 200 phosphenes there is only difference between 20 and 40 FOVs. In the case of recognition time (see Figure~\ref{fig8}\subref{fig8b}), there is difference between the two resolution for 40 and 60 FOVs. At the same time, there is difference between 40 and 60 FOVs for 500 phosphenes and between 20 and 40 FOVs for 200 phosphenes. Note that the data distribution for the condition 20 FOV and 200 resolution and the condition 40 FOV and 500 resolution are very similar, even though they are two totally different schemes. However, it makes sense because they have practically the same angular resolution (see Figure~\ref{fig9}).

Figure~\ref{fig9} shows the ratio of recognized objects and time recognition according to \emph{angular resolution} ($AR$), measured in phosphenes per radian. We perform a logarithmic regression for all data. The logarithmic regression equation for the \emph{object recognition rate} ($OR$) is $OR = -1.0345 + 0.4482\cdot \log(AR)$ with $R^{2} = 0.4031$ and $F-value = 201.3$. The positive coefficient indicates that as the angular resolution increases, the object recognition ratio tends to increase (see Figure~\ref{fig9}\subref{fig9a}). The logarithmic regression equation for the \emph{recognition time} ($RT$) is $RT = 83.14 - 18.78\cdot\log(AR)$ with $R^{2} = 0.2344$ and $F-value = 91.26$. The negative coefficient indicates that as the angular resolution increases, the recognition time tends to decrease (see Figure~\ref{fig9}\subref{fig9b}). The regression output shows that both, ratio of recognized object and the recognition time variables are statistically significant with $p-value<$ 0.05 (see Table~\ref{tab:table1}).

\section{Discussion}

\begin{figure}[t!]
\centering
\subfigure[]{\includegraphics[width = 3in]{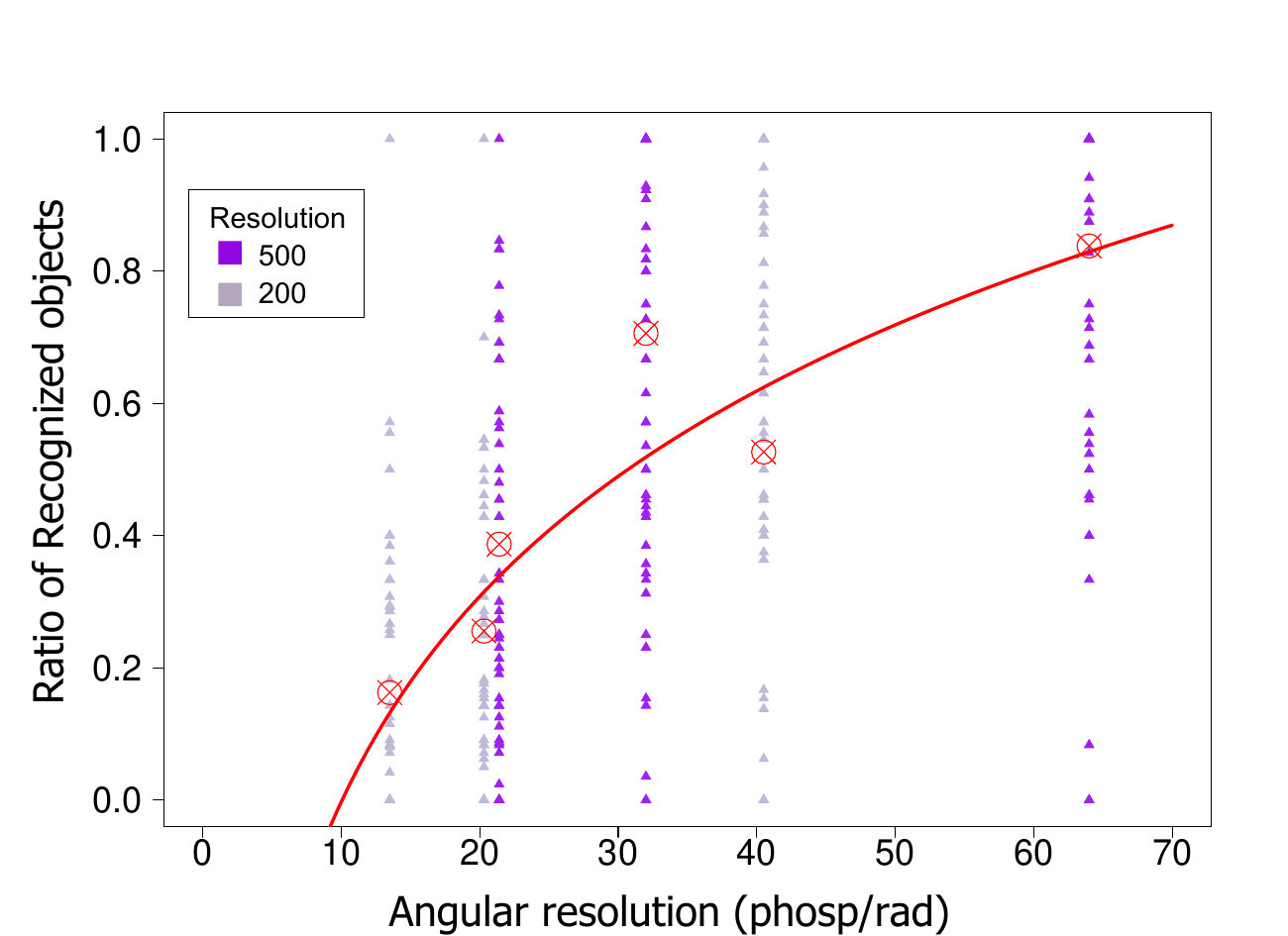}\label{fig9a}} 
\subfigure[]{\includegraphics[width = 3in]{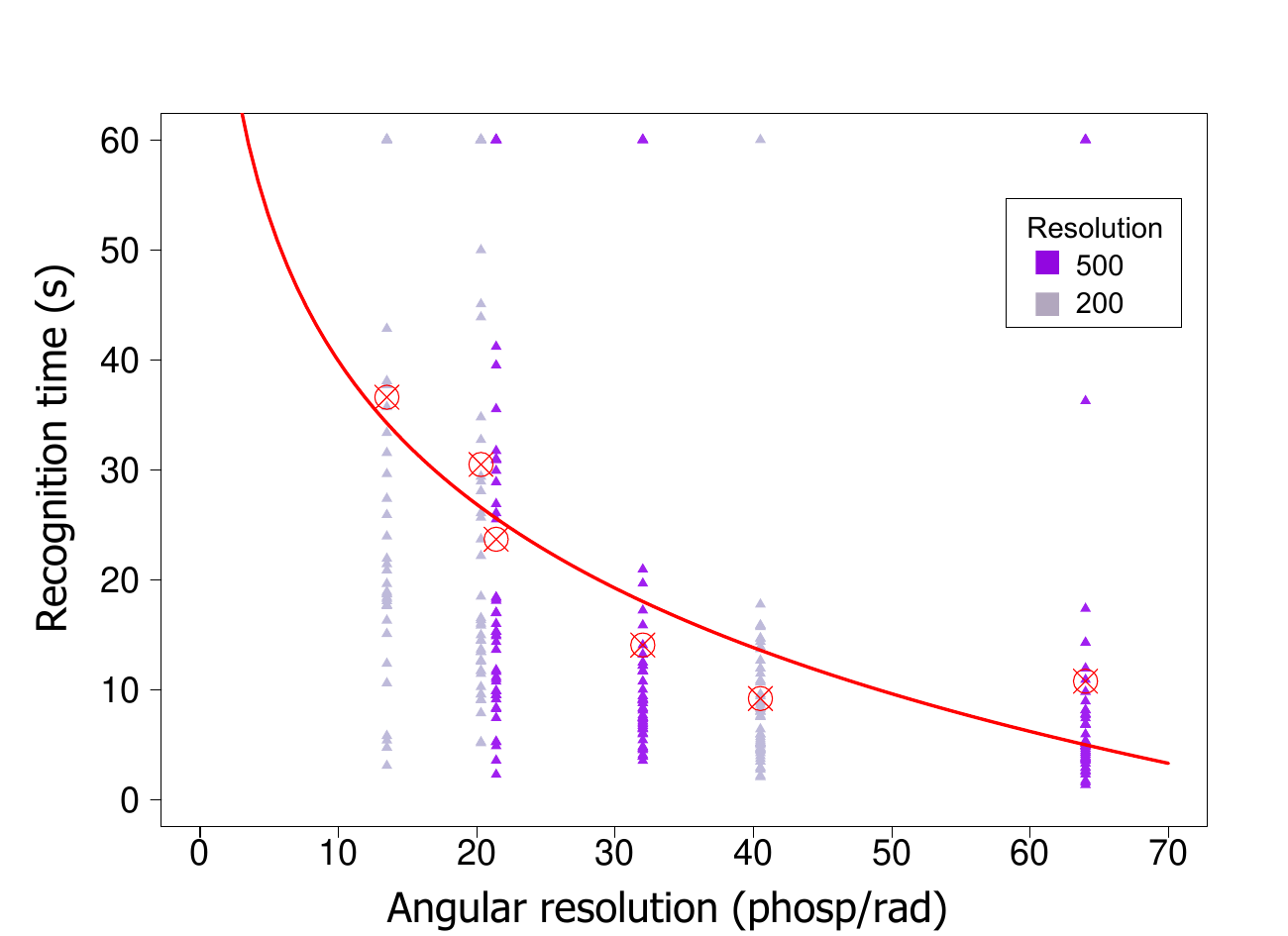}\label{fig9b}}
\caption{\textbf{Angular resolution.} \subref{fig9a} Performance of recognized objects according to angular resolution (phosphenes/radian) and \subref{fig9b} recognition time according to angular resolution. We perform a logarithmic regression for all data. The performance of recognized object tends to increase with angular resolution. Contrary, the recognition time tends to decrease with the angular resolution.}
\label{fig9}
\end{figure}

\begin{table}
\caption{\label{tab:table1} Mean value of the ratio of recognized object and recognition time for the six angular resolutions. Parameters for logarithmic regression and $R^{2}$, $F-value$ and $p-value$.}
\begin{indented}
\item[]\begin{tabular}{cccc}
\br
Angular resolution & Ratio of & Recognition time & $\#$ Results\\
(phosp/rad)  &  Recognized object & (s)  & $>60s$ \\
\mr
13.5 & 0.166 $\pm$ $0.058$ & 36.61 $\pm$ 5.77 & 20\\
20.3 & 0.197 $\pm$ $0.059$ & 30.50 $\pm$ 5.79 & 14\\
21.4 & 0.346 $\pm$ $0.078$ & 23.71 $\pm$ 4.91 & 7\\
32.0 & 0.638 $\pm$ $0.087$ & 14.10 $\pm$ 4.41 & 4\\
40.5 & 0.602 $\pm$ $0.083$ & 9.24 $\pm$ 2.35 & 1\\
64.0 & 0.807 $\pm$ $0.076$ & 10.81 $\pm$ 4.32 & 4\\ \hline
$intercept$ & $-1.0345$ & $83.14$ & $-$ \\
$slope$ & $0.4482$ & $-18.78$ & $-$ \\
$R^{2}$ & 0.4031 & 0.2344 & $-$ \\
$F-value$ & 201.3 & 91.26 & $-$ \\
$p-value$ & $***$ & $***$ & $-$ \\ 
\br
\end{tabular}
\end{indented}
\end{table}
\normalsize

Recent developments in the technology for retinal implants should allow to improve the quality of electrode arrays, in terms of features, such as response time, resolution or size \cite{choi2017human,ameri2009toward}. For example, Beauchamp et al. \cite{beauchamp2020dynamic} conducted a study with blind patients with cortical visual prostheses in which shapes were traced on the surface of the visual cortex by stimulating electrodes in dynamic sequence. The dynamic stimulation allowed rapid and accurate recognition of letter shapes predicted by the spatial map of the visual world of the brain. However, an open question for the prosthesis design is how to evaluate and predict the utility and functionality in terms of patient benefit with respect to design parameters \cite{bloch2019advances}. Several studies have incorporated  performance-based measures and questionnaires to understand the relation of visual parameters in the performance of everyday tasks in visually impaired subjects.

In this work, we study how certain features interact and affect perception. Intuitively, wider sensors should improve perception by allowing a larger field of view which can help detect motion and perform scanning \cite{ameri2009toward}. Our results seem to indicate that it is not the case for head scanning, if it comes at the expense of phosphene density. Similarly, He et al. \cite{he2019trade} investigated the effect of a wider FOV of a retinal prostheses on the performance of Argus II users in an object localization task using a thermal sensor. The results showed that users were able to find objects using the current $11^{\circ}$ $\times$ $18^{\circ}$ FOV with no zoom with higher precision and speed than when using a wider FOV by zooming out the sensor input. They also suggest that a higher spatial resolution may be preferred over a wider input FOV. This has demonstrated the importance of some parameters such as FOV, visual resolution and angular resolution \cite{haymes2002relationship}. Others’ work has also shown that the lack of visual information caused by the low resolution and the restriction in a large portion of the FOV can be compensated by scene scanning \cite{dobelle2000artificial,humayun2003visual}. During scene scanning, the relevant information of the environment is actively sought, quickly and efficiently. For instance, there is some evidence on humans that eye and head rotation facilitates the learning process to recognize simple objects \cite{dobelle2000artificial,brelen2005creating,veraart2003pattern,ehrlich2017head}.

Simulated prosthetic vision (SPV) can be used to estimate the visual requirements to deliver a sufficient visual resolution and FOV with a large statistical power and cost efficiency. It provides an opportunity for simulation-based research regarding the design of a functional retinal prosthesis, improvement of functional vision in low-phosphene-count devices, and also as a tool to search for image processing strategies to impart the most understandable prosthetic vision making experimentation with implanted patients unnecessary. In the work of He at al. \cite{he2019trade}, the FOV is modified at the external sensor by introducing a zoom lens, therefore changing the perceived scale. 
In contrast, thanks to the use of SPV we can actually change the properties of the electrode array, altering the FOV at the stimulus, maintaining the 1:1 scale ratio between perception and real world.

Visual scanning has also been studied in SPV showing that it has a positive impact on the task performance by allowing subjects to increase the visual information \cite{cha1992simulation}. Cai et al.\cite{cai2006prosthetic} noticed during experiments that head rotation also allowed subjects to expand their effective FOVs and rudimentary depth perception through parallax. Chen et al.\cite{chen2006psychophysics} also reported that maximizing visual information with scanning techniques improves visual acuity performance.

The theoretical visual acuity achievable by present-day retinal implants such as Argus II is $4^\circ$ with a FOV of $20^\circ$ across the diagonal \cite{stronks2014functional}. There have also been reported cases in Argus II clinical trials with $1.1^\circ$, well below the theoretical limit \cite{humayun2012interim}. This may be explained by effective scanning techniques, allowing subjects to temporally integrate percepts. Previous studies in SPV found out that a minimum angular resolution of 170 phosphenes/radian is needed for an acceptable accuracy in some tasks such as mobility and object recognition \cite{fornos2008simulation,sommerhalder2017prospects,dagnelie2007real,hayes2003visually}. However, this angular resolution is around three times more than the maximum angular resolution obtained in our experiment (64.0 phosphenes/radian).

We found that object recognition is well achieved with low resolution and restricted FOV. As can be seen in Figure~\ref{fig7}\subref{fig7a}, we obtained a significant improvement in the recognition performance as the FOV was reduced, for both resolutions. Besides, participants took less time to recognize objects with a narrower FOV (see Figure~\ref{fig7}\subref{fig7b}). This seems counterintuitive since with a narrower FOV the global reference of the scene is lost. However, the narrower the FOV, the higher the angular resolution and therefore the greater the image detail (higher frequencies). Contrary, the widest FOV allows to cover the widest area of the scene but it only allows to see the gist of the image (lower frequencies). 

Psychophysical and computational studies have shown that high and low spatial frequency provide different content from a scene: higher spatial frequencies contain fine information of image details and/or object boundaries, whereas lower frequencies preserve coarse blobs representing the gist of the scene \cite{kihara2010time, oliva2006building, oliva2005gist, oliva1997coarse}. In our experiments, the narrower FOV produces a higher phosphene density or angular resolution. By increasing the FOV (20, 40, 60 degrees), the angular resolution decreases: 40.5, 20.3 and 13.5 phosphenes/radian for 200 phosphenes and 64.0, 32.0 and 21.4 phosphenes/radian for 500 phosphenes, as can be seen in Figure~\ref{fig9}. For lower angular resolutions, there were more cases of responses from subjects that exceeded the 60 s limit in the experiment, which implies that the subjects need more time to recognize details in these conditions (see Table~\ref{tab:table1}). Our results suggest that it is better to see higher spatial frequencies of the scene, even if the global reference is lost due to the narrow FOV, since the subjects are able to holding back the global concept of the scene through visual scanning. Further, in more complex scenes such as low-contrast or low-luminance level, the time required to recognize the objects was higher and recognition performance decreased. In the same way, in those scenes with high contrast the subjects needed less time and the performance in the task increased. This fact has been demonstrated by Ehrlich et al.\cite{ehrlich2017head}, where they observed that as the contrast increases the perception of the scene increases.

One limitation of our study is that the environment and the user location were static. We found that the dynamic task of seeking objects did not benefit for a larger FOV. However, for more complex
tasks, such as navigation, where location must also be inferred and with potential collisions with dynamic objects, the benefits of a larger FOV could compensate for the reduced angular resolution. Hassan et al. \cite{hassan2007minimum} studied the minimum FOV size required for efficient navigation. Their results suggested that the size of the FOV required for safe navigation ranged between 10.9º and 32.1º but it was dependent on average image contrast. Furthermore, with a low resolution implant, rendering a complex visual scene, potentially dynamic, with no preprocessing leads to visual overcrowding, and does not provide sufficient visual cues to localize or navigate an unknown environment. Dealing with the same problem, we have developed a framework for perception and navigation where only the salient information of the scene is presented to the user, considering
critical aspects such as obstacle avoidance and using a head-mounted RGB-D camera to detect free-space, obstacles and scene direction in front of the user \cite{perez2017depth} or critical features to help identify the user location \cite{sanchez2020semantic}. One advantage of our experimental setup is that we could quickly generate navigation tasks and dynamic environments, while maintaining participant safety.

Virtual-reality (VR) systems have been widely used for experimentation \cite{denis2014simulated, denis2012simulated, chen2005visual, van2010simulating, vergnieux2017simplification}. They improve the quality of SPV tests since the experimentation environment is closer to the real-world. Some SPV research such as Denis et al. \cite{denis2014simulated} used a VR system with two videos cameras mounted on the front of the headset to capture the visual scene for text localization. In \cite{denis2012simulated}, the SPV was generated on a VR headset with a stereo camera that captured the scene in real-time for object recognition and localization. In similar studies, participants interacted with the VR system for different tasks such as visual acuity measurement \cite{chen2005visual} or wayfinding task \cite{van2010simulating}. Vergnieux et al. \cite{vergnieux2017simplification} used a virtual environment that was displayed via a VR system for navigation task evaluation. One advantage of VR systems is that they could also allow for further experimentation working with dynamic and complex environments, which could benefit from dynamic stimulation \cite{beauchamp2020dynamic}.

Despite these proposals, we need more realistic and easier to use experimentation environments to improve the quality of SPV tests. A complete immersive experience implies to consider rotations and translations in the virtual environment. This approach would require the development of a detailed 3D model of the scene, to track the user and a large space allowing the free movement of the user in the real world. Since the influence of visual scanning in the FOV is dominated by rotations we have considered a simpler option. The environment is represented by a 360 panoramic scene so that the head can be turned in all directions from the center of the scene and the subject can explore the entire scene. This approach has a set of advantages. Panoramic scenes are easy to obtain in any real scene. By contrast, modelling a complete 3D model of the environment is very expensive. Further, the rotational movement of the head can be easily obtained from the embedded inertial sensors of the headset avoiding external cameras for the translation tracking. The experiment can be performed sitting on a chair, avoiding collisions, the requirement of a very large space and facilitating repeatability. Furthermore, our SPV system allows replication of the experiment using many subjects with normal vision without being limited by the number of implanted subjects.

\section{Conclusions}

Effective retinal prosthesis design presents a challenge for engineers and neuroscientists. We have analyzed the influence of field of view with respect to resolution in retinal prostheses through a study with a novel simulated prosthetic vision setup: a virtual-reality system using panoramic scenes. Participants perceived phosphene images in an immersive head-mounted device. The task consisted on finding and identifying common objects with different field of view and number of phosphenes. Our results show that, for the same number of phosphenes, recognition accuracy and response time improved by reducing the field of view. In fact, angular resolution is major determinant for effective object recognition, being directly correlated to the accuracy and inversely correlated to the response time. However, it has also shown a diminishing return even for an angular resolution of less than 2.3 phosphenes per degree. Simulated prosthetic vision allows experiments with larger number of participants and simpler procedures than clinical studies with implanted patients. Our experimental setup relies on a consumer-level head-mounted display, public image databases and we have released the software needed to run the simulator, to facilitate replications and extensions. Our results indicate that the phosphene versus field-of-view trade-off for improvements to the angular resolution should prioritize the former to the latter.

\section*{Acknowledgments}

This work was supported by project RTI2018-096903-B-I00 (MINECO/FEDER, UE) and BES-2016-078426 (MINECO). The authors thank Maria Santos Villafranca for collaborating in the VR system development.

\section*{Ethical statement}

The research process was conducted according to the ethical recommendations of the Declaration of Helsinki. The research protocol used for this study is non-invasive, purely observational, with absolutely no-risk for any participant. There was no personal data collection or treatment and all subjects were volunteers. Subjects gave their informed written consent after explanation of the purpose of the study and possible consequences. The consent allowed the abandonment of the study at any time. All data were analyzed anonymously. The experiment was approved by the Aragon Autonomous Community Research Ethics Committee (CEICA).

\section*{Appendix A\label{sec:AppendixA}}
\appendix
\setcounter{section}{1}
\renewcommand{\theequation}{A.\arabic{equation}}
\renewcommand{\thesection}{A.\arabic{section}.}

\subsection*{Spherical image projection in the VR system}

Since the participants only rotate the head and body during the experiment, in our SPV system, we only consider rotation movements.
In the absence of translation, all the visual information of the environment can be encoded in a spherical image, i.e. an equirectangular panorama. 
Thus, if the rotation of the head is known, the image received by the subject through the screen can be virtually simulated from a panoramic image.

To achieve the effect of the stimulus image on the VR system we combine two different projections models: the spherical projection that models the projection on the panoramic image and the perspective projection that models the projection on the viewfinder screen.

In the Oculus VR system, the inertial measurement unit (IMU) collect the information of the head movement made by the user in form of quaternions like $q = a + bi + cj + dk$. This information is used to generate the rotation matrix which describes the basis of the absolute reference system used by the IMU with respect to the screen reference system as:

\begin{equation}
\label{eq:1}
R_{imu} = \left(\begin{array}{ccc}
1 - b^2 - c^2 & 2ab - 2cd & 2ac + 2bd\\
2ab + 2cd & 1 - 2a^2 - 2c^2 & 2cb + 2ad\\
2ac - 2bd & 2bc + 2ad & 1 - 2a^2 - 2b^2
\end{array}\right)
\quad
\end{equation}

This rotation matrix defines the head orientation and, therefore, it can be used to find the area of the panoramic scene that is currently being observed. Then, the selected area is projected on the screen. 

For the generation of the screen image, we calculate the ray corresponding to each phosphene on the screen as

\begin{equation}
\label{eq:2}
\vec{v} = R_{pan}^{T} R_{imu}^{T} K^{-1} u
\end{equation}

\noindent where $\vec{v} = \left(v_x,v_y,v_z\right)^{T}$ is the direction vector of the ray in the panorama reference system, $R_{imu}$ is the rotation matrix from the IMU data, $R_{pan}$ is the the basis of the panorama reference system with respect to the absolute reference system of the IMU and $u = (j,i,1)^{T}$ are the coordinates of each phosphene on the screen image. $K$ is the calibration matrix used in the perspective projection that models the projection in the screen.

The spherical coordinates of the ray $\vec{v}$, known as azimuth $\phi$ and elevation $\theta$ angles, and computed as:

\begin{equation}
\label{eq:8}
\phi = \mbox{atan2}(v_y,v_x)
\end{equation}

\begin{equation}
\label{eq:9}
\theta = \sin^{-1} \frac{v_z}{\sqrt{v_x^{2} + v_y^{2} + v_z^{2}}}
\end{equation}

\noindent and are related with the coordinates of the pixels in the panorama though the linear mapping,

\begin{equation}
\label{eq:10}
x_{pan} = x_{0} + \frac{\phi}{2\pi} W
\end{equation}

\begin{equation}
\label{eq:11}
y_{pan} = y_{0} + \frac{\theta}{\pi} H
\end{equation}

\noindent where $(x_0,y_0)$ is the center of the panoramic image, and $(W, H)$ are, respectively, the width and height of the panoramic image in pixels.

\section*{Appendix B\label{sec:AppendixB}}
\appendix
\setcounter{section}{2}
\renewcommand{\theequation}{B.\arabic{equation}}

\renewcommand{\thesection}{B.\arabic{section}.}

\subsection*{Phosphene generation}

In order to simulate the visual perception in prosthetic vision, our phosphenes generated are idealized representations of the percepts feasible in the current implants \cite{humayun2003visual, humayun2004clinical, weiland2003electrical, mahadevappa2005perceptual,richard2007chronic}. For the generation of phosphenes we start from the final image that is shown by the viewfinder (see Appendix A). First, we calculated the number of phosphenes to represent (200 or 500 phosphenes) depending on image size. Once the size of the phosphenes is known, the position of each phosphene is calculated from the image size, the viewing radius, the size of the phosphenes, the calibration array, and the center point of the image. The result is a array of rays associated with each phosphene. We use this array of rays in real-time, the size of which is much smaller than if we go through the viewer image pixel by pixel.

To generate the phosphenes we use the position and size previously calculated. Different patterns of phosphenes have been used for simulation such as square \cite{chai2007recognition}, hexagonal \cite{hallum2003filtering} or irregular pattern \cite{cai2006prosthetic}. One way to specify a pattern configuration is by changing the distance between phosphenes. However, the most commonly phosphene pattern adopted by SPV in the literature is square array \cite{chen2009simulating}. Similarly to many SPV studies \cite{chen2009simulating}, phosphenes are approximated as single channel (green) circular dots with a Gaussian luminance profile. We use a single color channel to avoid the effect of chromatic aberration of the Oculus display, and we select the green channel because it is the color channel best perceived by the human eye under normal conditions. Note that in terms of information provided to the user, this is equivalent to use a single gray-scale channel. The luminance profile of each phosphene has maximum intensity at the center and gradually decays to the periphery, following an unnormalized Gaussian function $G(x,y)$ defined in Equation~(\ref{eq:gaussianEq}).

\begin{equation}
\label{eq:gaussianEq}
G(x,y) \propto \exp\left\{\frac{1}{2} \frac{(x-\mu_x)^2+(y-\mu_y)^2}{\sigma^2 \cdot \mathcal{L}^2(x,y)}\right\}
\end{equation}
where the radial (isotropic) standard deviation of the Gaussian profile is $\sigma \cdot \mathcal{L}(x,y)$ with $\mathcal{L}(x,y)$ being the normalized intensity. 

\begin{equation}
\label{eq:intensityEq}
\mathcal{L}(x,y) = \frac{\mathcal{I}(x,y)}{\max(l)}
\end{equation}

The intensity of a phosphene is a function \(f:{\rm I\!R}^n\rightarrow{\rm I\!R}\) of the intensity of the pixels of the input image, usually the average of the a small window. Each intensity $(\mathcal{I})$ is quantified to each individual phosphene dynamic range $\mathcal{I}(x,y) \in \{0, \ldots, l-1\}$ where $l$ is the number of intensity levels of the phosphene dynamic range. 

Then, the final pixel value for the simulated image is:
\begin{equation}
\label{eq:brightnessEq}
P(x,y) = \mathcal{L}(x,y) \cdot G(x,y)
\end{equation}
Thus, both the size and brightness are directly proportional to the quantified sampled pixel intensities $\mathcal{L}(x,y)$. 

\section*{References}

\bibliographystyle{iopart-num}
\bibliography{mybibfile}

\end{document}


\maketitle

\section{Phosphene generation}

In order to simulate the visual perception in prosthetic vision, our phosphenes generated are idealized representations of the percepts feasible in the current implants \cite{humayun2003visual, humayun2004clinical, weiland2003electrical, mahadevappa2005perceptual,richard2007chronic}. For the generation of phosphenes we start from the final image that is shown by the viewfinder (see Appendix B). First, we calculated the number of phosphenes to represent (200 or 500 phosphenes) depending on image size. Once the size of the phosphenes is known, the position of each phosphene is calculated from the image size, the viewing radius, the size of the phosphenes, the calibration array, and the center point of the image. The result is a array of rays associated with each phosphene. We use this array of rays in real-time, the size of which is much smaller than if we go through the viewer image pixel by pixel.

To generate the phosphenes we use the position and size previously calculated. Different patterns of phosphenes have been used for simulation such as square \cite{chai2007recognition}, hexagonal \cite{hallum2003filtering} or irregular pattern \cite{cai2006prosthetic}. One way to specify a pattern configuration is by changing the distance between phosphenes. However, the most commonly phosphene pattern adopted by SPV in the literature is square array \cite{chen2009simulating}. Similarly to many SPV studies \cite{chen2009simulating}, phosphenes are approximated as green circular dots with a Gaussian luminance profile. The luminance profile of each phosphene has maximum intensity at the center and gradually decays to the periphery, following an unnormalized Gaussian function $G(x,y)$ defined in Equation~(\ref{eq:gaussianEq}).

\begin{eqnarray}
\label{eq:gaussianEq}
G(x,y) \propto \exp\left\{\frac{(x-\mu_x)^2+(y-\mu_y)^2}{2\sigma^2lum}\right\}
\end{eqnarray}

The intensity of a phosphene is a function \(f:{\rm I\!R}^n\rightarrow{\rm I\!R}\) of the intensity of the pixels. Each sampled pixel intensity $(i)$ is quantified to each individual phosphene's dynamic range as:

\begin{eqnarray}
\label{eq:intensityEq}
lum(x,y) = \frac{i}{max(l)}
\end{eqnarray}

In Equation~(\ref{eq:intensityEq}), $l$ is the number of gray levels intensity of the phosphene. The size and brightness are directly proportional to the quantified sampled pixel intensities. Then, the phosphene profile $P(x,y)$ is applied to every phosphene resulting in the final image:

\begin{eqnarray}
\label{eq:brightnessEq}
P(x,y) = lum\cdot G(x,y)
\end{eqnarray}

where $P(x,y)$ represents the pixel value at the coordinates $(x,y)$ of the stimulus image (Equation~(\ref{eq:brightnessEq})).

\bibliographystyle{iopart-num}
\bibliography{mybibfile}


\maketitle

\section{Generation of the stimulus image in the VR system}
\label{section:1} 

In our SPV system, subjects only use rotation, not translation movements since they only rotate the head and body during the experiment. In the absence of translation, all the visual information of the environment can be encoded in a spherical image. Thus, the perspective image received by the subject can be virtually simulated from a panoramic image.

To achieve the effect of the stimulus image on the VR system we combine two different projections models: the spherical projection that models the projection on the panoramic image and the perspective projection that models the projection on the viewfinder screen.

In the Oculus VR system, the inertial measurement unit (IMU) or sensors collect the information of the head movement made by the user in form of quaterniums like $q = a + bi + cj + dk$. This information is used to generate the rotation matrix as:

\begin{equation}
\label{eq:1}
R = \begin{pmatrix}
1 - b^2 - c^2 & 2ab - 2cd & 2ac + 2bd\\
2ab + 2cd & 1 - 2a^2 - 2c^2 & 2cb + 2ad\\
2ac - 2bd & 2bc + 2ad & 1 - 2a^2 - 2b^2
\end{pmatrix}
\quad
\end{equation}

The rotation matrix allows choosing the area of the panoramic scene that is currently being observed. Then, the selected area is projected on the viewfinder as a perspective image. 

For the generation of the perspective image, we calculate the corresponding ray for each point of the viewfinder image as:

\begin{eqnarray}
\label{eq:2}
\vec{v} = R^{-1} K^{-1} u
\end{eqnarray}

where $\vec{v}$ is the direction vector of the ray, $R$ is the rotation matrix and $u = (j,i,1)^{T}$ is one point of the viewfinder image. $K$ is the calibration matrix used in the perspective projection that models the projection in the viewfinder.

Also, we use an additional variable called $R_{cam}$ to set the absolute reference system:

\begin{eqnarray}
\label{eq:4}
\vec{v}_{s} = R_{cam} R^{-1} K^{-1} P
\end{eqnarray}

\begin{figure}[h]
\centering
\includegraphics[width = 2in]{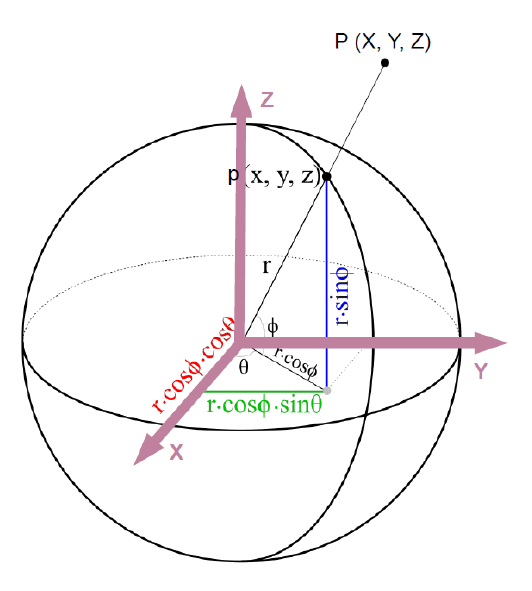}
\caption{\textbf{Spherical projection model}. Unit sphere model used to project 3D points into 2D image.}
\label{Figure1}
\end{figure}

The $\vec{v}_{s}$ ray corresponds to the 3D point $P(X,Y,Z)$ of the spherical projection, as can be seen in Figure~\ref{Figure1}.

Using the spherical projection model, each point is projected into the sphere surface. For that, we converted the sphere coordinates

\begin{eqnarray}
\label{eq:5}
X = \sqrt{ X^{2} + Y^{2} + Z^{2}}  \cos{\theta}  \cos{\phi}
\end{eqnarray}

\begin{eqnarray}
\label{eq:6}
Y = \sqrt{ X^{2} + Y^{2} + Z^{2}}  \cos{\theta}  \sin{\phi}
\end{eqnarray}

\begin{eqnarray}
\label{eq:7}
Z = \sqrt{ X^{2} + Y^{2} + Z^{2}}  \sin{\theta}
\end{eqnarray}

to azimuth $\phi$ and latitude $\theta$ angles as:

\begin{eqnarray}
\label{eq:8}
\phi = \atantwo(Y,X)
\end{eqnarray}

\begin{eqnarray}
\label{eq:9}
\theta = \sin^{-1} \frac{Z}{\sqrt{X^{2} + Y^{2} + Z^{2}}}
\end{eqnarray}

And then, to the 2D image coordinates:

\begin{eqnarray}
\label{eq:10}
x_{ij} = x_{0} + \frac{\phi}{\pi} \frac{W}{2}
\end{eqnarray}

\begin{eqnarray}
\label{eq:11}
y_{ij} = y_{0} + \frac{\theta}{\pi} H
\end{eqnarray}

where $W$ and $H$ are, respectively, the width and height of the panoramic image in pixels.
